\documentstyle[prb,aps,multicol,epsf]{revtex}

\def\vk{{\bf k}}   \def\vp{{\bf p}}

\def\be{\begin{equation}}  \def\ee{\end{equation}}
\def\bea{\begin{eqnarray}} \def\eea{\end{eqnarray}}

\def\Bildeins{\epsfbox{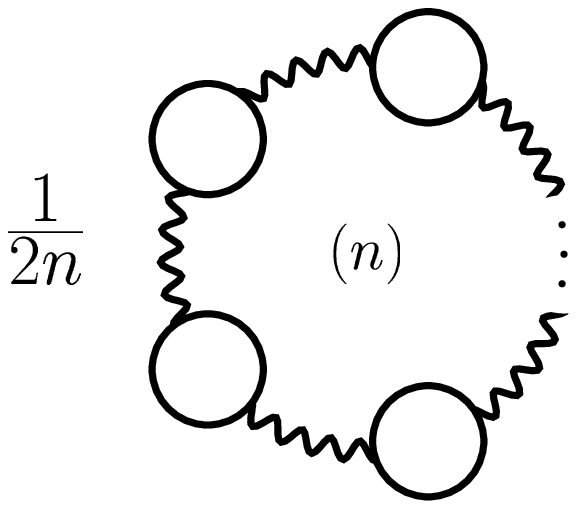}}
\def\Bildzwei{\epsfbox{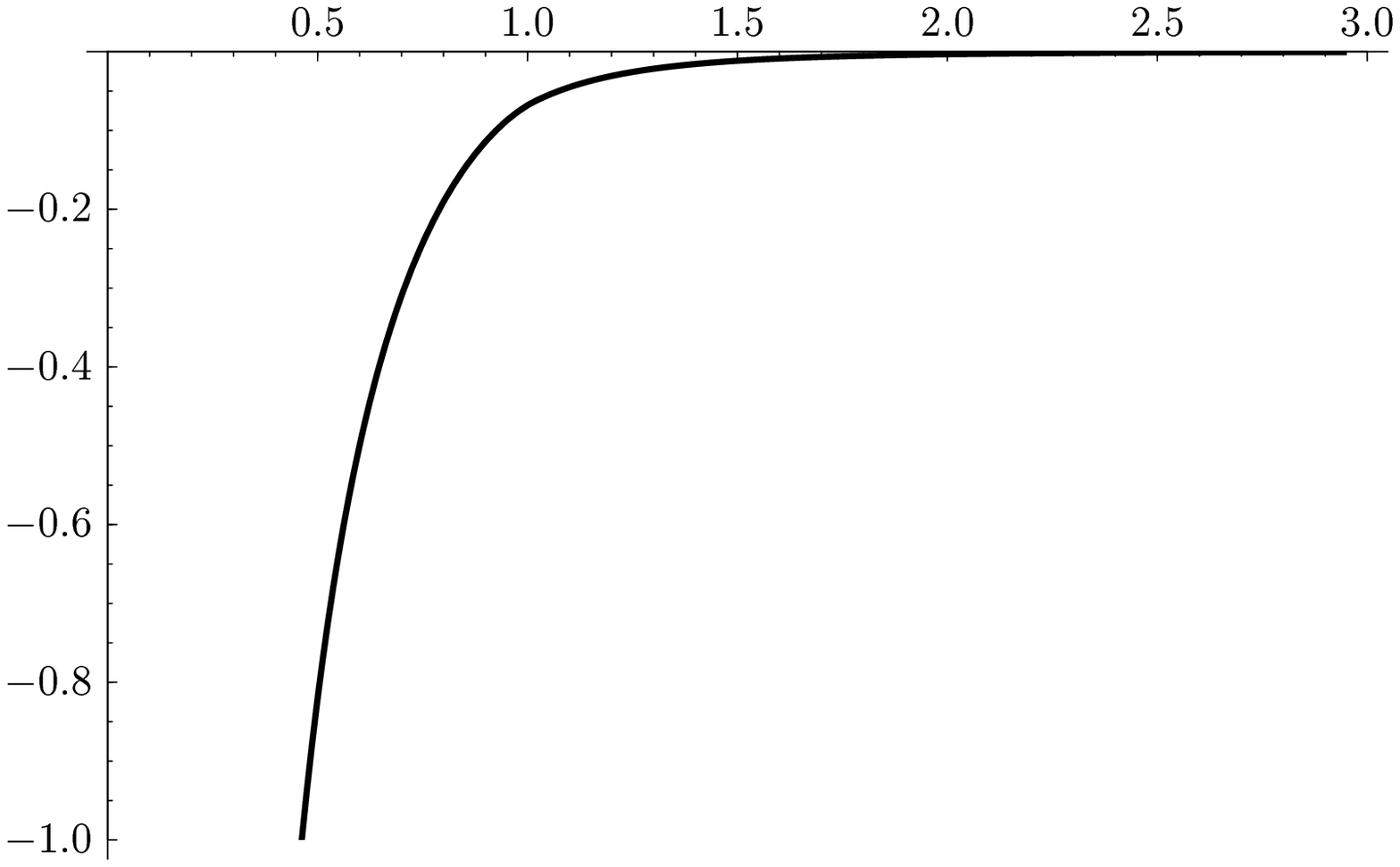}}
\def\Bilddrei{\epsfbox{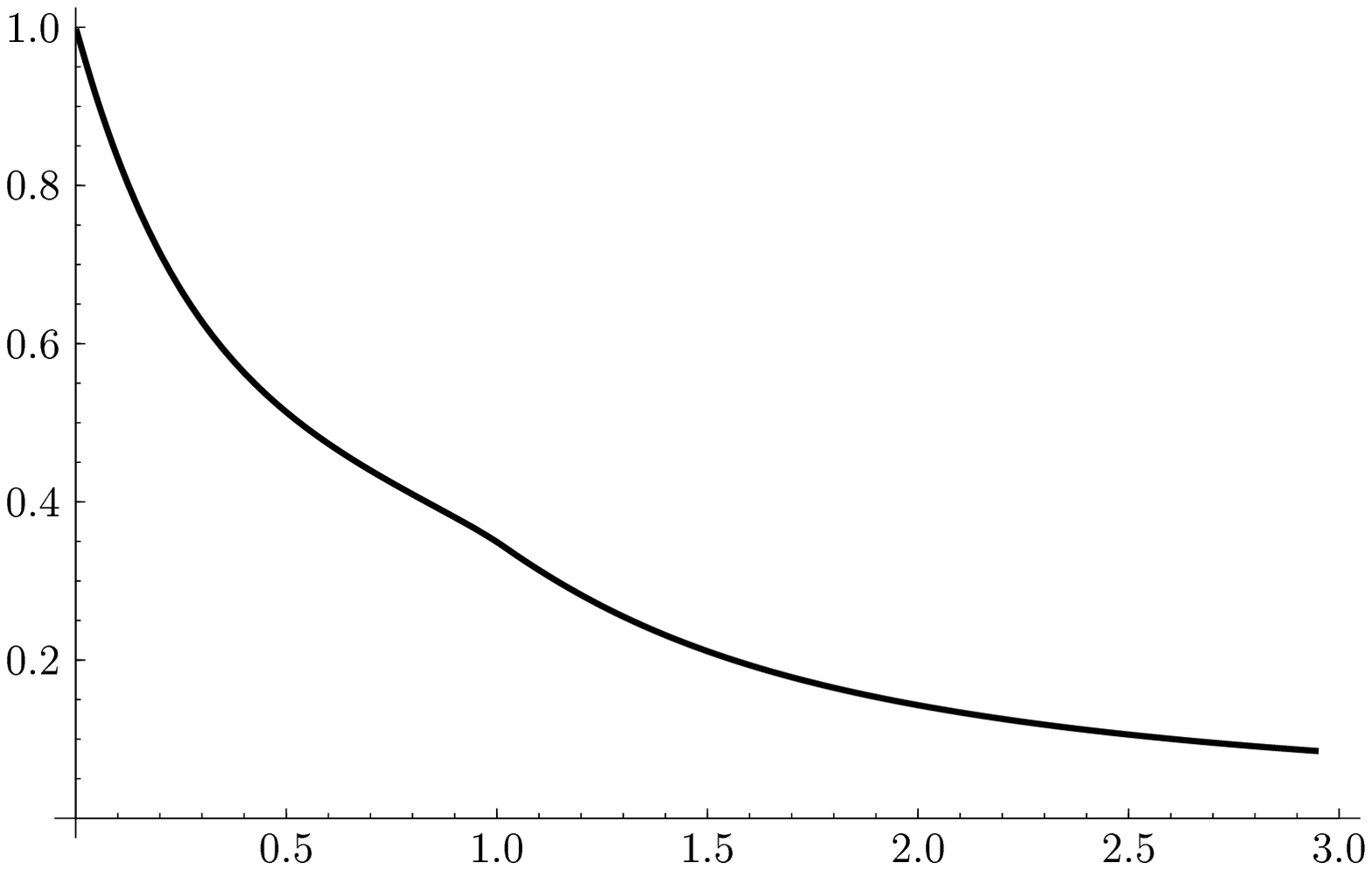}}
\def\Bildvier{\epsfbox{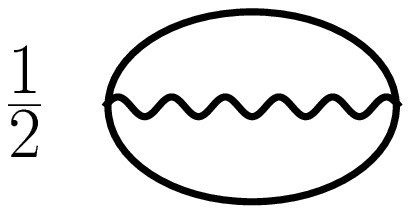}}
\def\Bildfuenf{\epsfbox{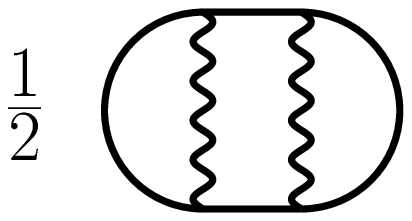}}
\def\Bildsechs{\epsfbox{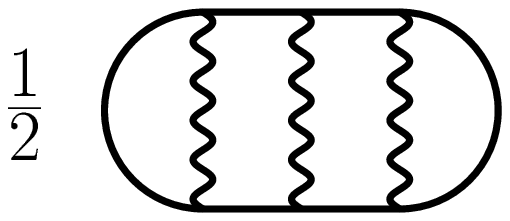}}
\def\Bildsieben{\epsfbox{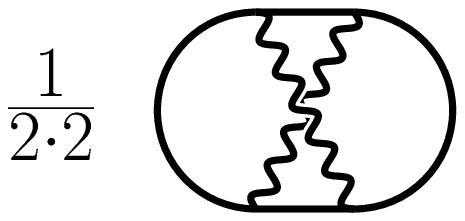}}
\def\Bildacht{\epsfbox{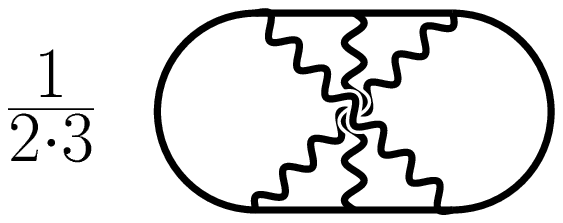}}

\begin{document}
\draft

\title{Random--phase approximation for the grand--canonical potential of
composite fermions in the half--filled lowest Landau level }
\author{J. Dietel$^{a}$, Th. Koschny$^{a}$, W. Apel$^{b}$, and W. Weller$^{a}$}
\address{
$^{a}$ Institut f\"ur Theoretische Physik, Universit\"at Leipzig,\\
Augustusplatz 10, D 04109 Leipzig, Germany \\
$^{b}$ Physikalisch-Technische Bundesanstalt,\\
Bundesallee 100, D 38116 Braunschweig, Germany\\
}

\date{\today}
\maketitle

\begin{abstract}
We reconsider the theory of the half--filled lowest Landau level
using the Chern--Simons formulation and study the grand--canonical
potential in the random--phase approximation (RPA).
Calculating the unperturbed response functions for current-- and
charge--density exactly, without any expansion
with respect to frequency or wave vector, we find that
the integral for the ground--state energy converges rapidly
(algebraically) at large wave vectors $k$,
but exhibits a logarithmic divergence at small $k$.
This divergence originates in the $k^{-2}$ singularity of
the Chern--Simons interaction and it is already present
in lowest--order perturbation theory.
A similar divergence appears in the chemical potential.
Beyond the RPA, we identify diagrams for the grand--canonical potential
(ladder--type, maximally crossed, or a combination of both)
which diverge with powers of the logarithm.
We expand our result for the RPA ground--state energy in the strength
of the Coulomb interaction.
The linear term is finite and its value compares well with
numerical simulations of interacting electrons in the lowest Landau level.
\end{abstract}


\begin{multicols}{2}
\narrowtext

\section{Introduction}
The combination of an electronic interaction and a strong magnetic field
in a two--dimensional electron system yields a rich variety of phases.
These are best classified by the filling factor $\nu$, which is the electron
density divided by the density of a completely filled Landau level.
In the case of $\nu \cong 1/2$, the behavior of the system resembles that of
a Fermi liquid in the absence of a magnetic field, or at small magnetic
fields, and over the past years, a intriguing picture has emerged :
at $\nu =1/2$, each electron combines with two flux quanta of the magnetic
field to form a composite fermion; these composite fermions then move
in an effective magnetic field which is zero on the average.
The interpretation of many experiments supports this picture.
Here, we mention transport experiments with quantum (anti--) dots
\cite{KSPBW93}, in which features of the resistivity are related to
closed loops of the composite fermions around the dots,
and also focusing experiments \cite{SWBLvKFKGW96};
more references to recent work can be found in
\cite{SDKTPBW94,Heiman,S97,Das-Sarma-Pinczuk}.

Theoretically, the picture was developed in the pioneering work of
Halperin, Lee, and Read \cite{HLR93}.
They formulated the Hamiltonian in terms of Chern--Simons (CS)
transformed electrons and studied within the random--phase
approximation (RPA) many physical quantities.
Most prominent among these is the effective mass of the
composite fermions which they found to diverge at the Fermi surface.
This problem of the effective mass was again taken up by Stern and
Halperin \cite{SH95} who calculated the effective mass and also
the quasiparticle interaction function of Landau's Fermi--liquid theory,
see also \cite{SSH96,S96}.
Recently, Kopietz and Castilla \cite{KC97} used the method of
higher--dimensional bosonization to extend the analysis beyond the RPA.
Thus treating the infrared singularities, they found that the vertex
corrections render the quasiparticle mass finite.
But the composite fermion spectrum should scale with the strength
of the Coulomb repulsion.
Most recently, Shankar and Murthy \cite{SM97} proposed, within the CS approach,
a new parametrization which allows to separate the contributions
of the CS particles from those of the magneto--plasmon oscillators in the
Hamiltonian.
After they restrict the number of the magneto--plasmon oscillators
to the number of electrons in the original Hamiltonian, they get a finite
quasiparticle mass which scales with the inverse of the strength of the
Coulomb repulsion.

The CS formulation of interacting electrons, with a density such
that the filling factor of a Landau level is $\nu \cong 1/2$,
nicely realizes the concept of composite fermions,
in accordance with the experimental observations.
The price to be paid for using the CS transformation is that
the non--interacting system of electrons in a strong magnetic field,
after the transformation, becomes a higly complicated,
interacting problem -- formulated in terms of the CS field.
The CS interaction diverges for small wave vector $k$ as $1/k^2$.
Therefore, the singular diagrams in leading order need to be resummed;
these diagramms are well known as the RPA of the jellium problem in $d=3$.
But there is no small parameter in the problem of non--interacting electrons
in a strong magnetic field and there is no reason to expect that the
RPA diagrams already reproduce the correct result.
In fact, as we shall see, the RPA is even plagued with a marginal divergence
in the energy.
The understanding, and even the evaluation of the RPA, still seem to be
incomplete:

 Is it necessary to introduce an ultraviolet cut off
in the momentum integrals, in order to restrict the number of the
magneto--plasmon oscillators, or is that already inherent in the theory?
After all, the introduction of the cut off seems to be crucial for
the conclusions in the work of Shankar and Murthy.
The simplest quantity for a study of this question is the
ground--state energy.
Its evaluation in RPA should be straight forward.
But all approaches, of which we are aware, concentrate on the behavior at
small wave vector.
Therefore, we reconsider the RPA and calculate the expression for the
energy, refraining from any approximations in the bare response functions.

 There is a second motivation for studying the energy in RPA.
For physical reasons, it is clear that the exact energy per particle
should be expandable in the strength of the Coulomb interaction \cite{WDK97}.
The lowest--order term is exactly known, of course, it is the energy of the
unperturbed lowest Landau level; still, one can not expect to obtain
that result (or the corresponding degeneracy) correctly within the RPA.
Given the lowest order, we can proceed calculating -- in RPA -- the next,
the term linear in the strength of the Coulomb interaction.
Within a restriction to wave functions of the lowest Landau level,
there are no corrections beyond this first order.
The contribution of the Coulomb interaction to the total energy
seems not to be measurable;
still, there are numerical simulations of interacting electrons in the
lowest Landau level by Morf and d'Ambrumenil \cite{MA95} and
Girlich \cite{Girlich98} which are well suited for
a comparison with our analytical result.
We want to add that such an expansion with respect to the Coulomb interaction
should be useful also for the calculation of other quantities, as
the self energy of the composite fermions, because there 
the scale is set by the Coulomb interaction.

 The main problem of the RPA, which does not exist for
interacting electrons at zero magnetic field in $d=3$,
remains the logarithmic infrared divergencies caused by the $1/k^2$
behavior of the CS interaction at small wave vector.
Such a divergence already appears in the lowest order in the RPA.
Beyond the RPA, there are diagrams showing divergencies
with powers of the logarithm.

The outline of this work is as follows.
First, we introduce the model in its CS formulation.
Then, we calculate ground--state energy and chemical potential from the
grand--canonical potential in RPA using the exact bare response functions.
Further, we calculate the first term of the expansion of the energy in the
electron--electron (Coulomb) interaction.
Finally, we discuss the logarithmically divergent diagrams beyond the RPA.

\section{Chern--Simons description}
We consider interacting (spin aligned) electrons moving in two dimensions in a
strong magnetic field $B$ in the negative $z$--direction transverse to the
system.
The electronic density of the system is chosen such that the lowest
Landau level of a non--interacting system is filled to a fraction
of $\nu = 1/\tilde \phi$ where $\tilde \phi$ is an even integer.
We are mainly interested in $\tilde \phi = 2$.
Following the work of Halperin, Lee, and Read \cite{HLR93},
we describe the Hamiltonian in terms of Chern--Simons (CS) fermions or
composite fermions which
consist of the original electrons and $\tilde \phi$ attached flux quanta :
\bea
H = &&\sum_{\vk} \; (\epsilon_{k} - \mu) \; a^{\dagger}_{\vk} a^{ }_{\vk}
    \nonumber \\
  &&+ \frac{1}{2F}  \sum_{ {\vk_1, \vk_2} \atop {\vk \neq 0}} \;
     W_{ \vk ;\, \vk_1, \vk_2 } \;
     a^{\dagger}_{\vk_1} a^{\dagger}_{\vk_2 -\vk}
     a^{}_{\vk_2} a^{}_{\vk_1 - \vk}  \;\;.         \label{H}
\eea
Here, $a^{\dagger}_{\vk}$ creates (and $a^{ }_{\vk}$ annihilates)
a CS fermion with wave vector $\vk$.
Since we consider only cases where the filling factor is given by
$1/\tilde \phi$,
the sum of the external magnetic field and the mean CS field, $\Delta B$,
vanishes (compare Ref.~\cite{HLR93})
and $\epsilon_{k}$ is given by the free spectrum
$\epsilon_{k} = k^2/(2m_b)$, where $m_b$ is the electron band mass;
we use $\hbar =1$, but restore $\hbar$ in final formulae.
The chemical potential is denoted by $\mu$.
Because of $\Delta B = 0$, we can use periodic boundary conditions;
$F$ is the area of the sample.
The interaction has contributions from the fluctuations of the
CS vector potential and from the original Coulomb interaction
of the electrons, $V(r) = \epsilon^2 / r$, where
$\epsilon^2 = e^2/(4\pi \varepsilon_r \varepsilon_0)$ and $\varepsilon_r$ is
the dielectric constant.
Collecting all these contributions, we get
\be
     W_{ \vk ;\, \vk_1, \vk_2 } =
           W(k) + \frac{2\pi \tilde \phi}{m_b} \; i \left( ( \vk_1 - \vk_2 ) 
     \times \frac{\vk}{k^2} \right) {\bf e}_z  \; ;
\label{W}
\ee
${\bf e}_z$ denotes the unit vector in $z$ direction, orthogonal
to the system.
Here, the last term comes from the term linear in the CS vector potential.
The quadratic term is contained in $K(k)$ in
\bea
   W(k) &=& K(k) + V(k) \;,  \\
   K(k) &=&  \frac{(2\pi \tilde \phi)^2}{m_b} \, \frac{N}{F} \,
             \frac{1}{k^2} \;, \nonumber \\
   V(k) &=& \frac{2\pi \epsilon^2}{k} \;. \nonumber
\eea
Rewriting the original Hamiltonian in terms of CS operators $a^{\dagger}$,
$a$ as in the Hamiltonian (\ref{H}), we have made an approximation
which is already implicit in the treatment of Ref.~\cite{HLR93}~:
In the term quadratic in the CS vector potential, we have replaced a pair
of operators, $a^{\dagger}a$, by its average and neglected thus the
third order fluctuations in the density, which would lead to three particle
interactions, i.e., sixth order in $a^{\dagger}$, $a$.
This average leads to the appearance of the density $N/F$ in
the CS interaction $K(k)$; $N$ is the mean electron number,
the CS particle number.
For the filling factors $\nu$ which we consider here, we have
\be
    \frac{N}{F} = \frac{1}{2\pi l_B^2} \; \frac{1}{\tilde \phi } \;.
\label{filling}
\ee
$l_B$ is the magnetic length, $l_B^{-2} = |eB|/\hbar $.
The Hamiltonian (\ref{H}) is the starting point for a standard perturbation
theoretical (diagrammatic) treatment.
In the following, we will limit ourselves to a study of the
grand--canonical potential within the RPA.
This will enable us to derive explicit results for the ground--state energy
and the chemical potential.

\section{Grand--canonical potential}
The calculation of the grand--canonical potential $\Omega (T, F, \mu)$
in the RPA is described in text books \cite{Mahan}.
The only unusual feature in the present case is the (linear) dependence of
the interaction $W_{ \vk ;\, \vk_1, \vk_2 }$ on $\vk_1 - \vk_2$.
Thus, one has to consider two kinds of vertices in the perturbation theory,
a density vertex and a current vertex,
and the RPA equation becomes a matrix equation as stressed already in
Ref.~\cite{HLR93}.
The grand--canonical potential in the RPA, $\Omega_{\mbox{\scriptsize RPA}}$,
is given by the series of ring diagrams:
\begin{figure}[htb]
\setlength{\epsfxsize}{0.4\columnwidth}
\centerline{\Bildeins}
\label{omega}
\end{figure}
\be
\label{ring}
 \ = \ - \beta \Omega_{\mbox{\scriptsize RPA}}^{(n)} (T, F, \mu)
= {\cal O} \left( \Big[\frac{1}{\eta^2} \Big]^{n-1} \right) \;.
\ee
Here, we indicate the order of the leading singularity as the
infrared wave vector cutoff, $\eta \propto 1/\sqrt{F}$, decreases.
For $n=1$, this is a logarithmic singularity.
The singularity orginates from the divergence of the CS interaction
$K(k)$ at small wave vector.
The symmetry factor, $1/(2n)$, is explicitely written in the diagram.
We use the Matsubara technique and write the result as
\bea
&& \Omega_{\mbox{\scriptsize RPA}}(T,F,\mu) = \nonumber \\
&& - \frac{1}{\beta} \sum_{\vk} \ln \left\{ 1 + e^{-\beta (\epsilon_{k}-\mu)}
\right\} \nonumber \\
&& + \frac{1}{2\beta}  \sum_{\vk \neq 0} \sum_{\omega} \ln
 \left\{ 1 - \Pi_{00}(k, \omega)
             \left[ W(k) + \Pi_{11}(k, \omega) \right] \right\} \; .
\label{omRPA}
\eea
Here, the first term is the grand--canonical potential of the free
CS fermions.
The second term is the contribution of the magneto--plasmon oscillators.
In (\ref{omRPA}), the sum is on Matsubara frequencies of Bose type
($\omega = 2\pi n /\beta$, $n$ integer) and $\beta = 1/(k_B T)$.
The response functions of the unperturbed system, $\Pi_{00}$ and
$\Pi_{11}$, are defined and studied in the next section.

\subsection{Unperturbed response functions}
As explained in the introduction, our aim is to evaluate 
ground--state energy and chemical potential within the RPA
without additional approximations.
Therefore, we need to calculate the unperturbed response functions
exactly.
They are given by a one--loop integral involving
two unperturbed single--particle Greens functions.
The density--density response function (polarization part), $\Pi_{00}$,
was calculated for the jellium problem in $d=3$ by Lindhard \cite{L54}.
In the present case of $d=2$, fortunately, both response functions
can again be calculated analytically.
After performing the sum over Matsubara frequencies (of Fermi type),
we get
\bea
 \Pi_{00}(k, \omega) =&& \int \frac{d^2 p}{(2\pi)^2} \;
      \frac{e^{-i\omega 0^+} \; n_F(\epsilon_{p} - \mu) }
      {-i\omega
      + \epsilon_{p} - \epsilon_{|\vp - \vk|} }
   \nonumber \\
     && + ( \omega \rightarrow - \omega )
\eea
and
\bea
 \Pi_{11}(k, \omega) =
 && (\frac{2\pi \tilde \phi}{m_b})^2 \int \frac{d^2 p}{(2\pi)^2} \;
   \frac{(\vp \times \vk)^2}{k^4}
   \nonumber \\
   && \frac{e^{-i\omega 0^+} \; n_F(\epsilon_{p} - \mu) }
      {-i\omega
      + \epsilon_{p} - \epsilon_{|\vp - \vk|} }
    \nonumber \\
     && + ( \omega \rightarrow - \omega ) \; .
\eea
Here, $n_F(x)$ is the Fermi function.
The functions $\Pi_{00}$ and $\Pi_{11}$ depend only on the absolute value
of the wave vector $\vk$.
Note that in order to simplify our formulae, we absorbed a prefactor in
$\Pi_{11}$.
In the limit of zero temperature, when $n_F$ becomes the step function,
it is a tedious, but straight forward task to perform the remaining integrals
exactly without any approximative expansion in $k$ or $\omega$.
The zero--temperature results are
\be
 \Pi_{00}(k, \omega) = - \frac{m_b}{4\pi} \; e^{-i\omega 0^+} \;
     z( \frac{\omega}{\epsilon_{k}}, \frac{k}{2k_F})
     + ( \omega \rightarrow - \omega )
\ee
and
\bea
 \Pi_{11}(k, \omega) =
   && \frac{\pi {\tilde \phi}^2}{48 m_b} \; e^{-i\omega 0^+}
 \nonumber \\
    && \left( z( \frac{\omega}{\epsilon_{k}}, \frac{k}{2k_F})^3
     - 3 (\frac{2k_F}{k})^2 \,
         z( \frac{\omega}{\epsilon_{k}}, \frac{k}{2k_F})  \right)
     \nonumber \\
    && + ( \omega \rightarrow - \omega )  \;,
\eea
where $k_F$ is the Fermi wave vector defined by $\mu = k_F^2/(2m_b)$.
We use the abbreviation
\be
\label{z}
  z( x, u ) = 1 + ix - \sqrt{ (1+ix)^2 - u^{-2} } \;,
\ee
and we define the square root as analytical continuation from positive values
of the root for positive radicands.
Our results, the zero--temperature response functions $\Pi_{00}$
and $\Pi_{11}$, are correct for arbitrary large or small values
of $\omega /\epsilon_{k}$ and $k/(2k_F)$.
We still keep the convergence factor $\exp(-i\omega 0^+)$, because that
will be crucial in the following, where we have to perform the frequency
integral over $\Pi_{00}$ ($\sim 1/\omega$ for large $\omega$).

\subsection{Ground--state energy}
The ground--state energy coincides with the grand--canonical potential for
$T \rightarrow 0$ after adding the term $N\!\mu$.
Using the expression (\ref{omRPA}), we get
(here, $n_F(x)$ denotes the Fermi function at zero temperature)
\bea
&& E_{\mbox{\scriptsize RPA}}
=  \sum_{\vk} \; n_F(\epsilon_{k}-\mu) \; \epsilon_{k} \nonumber \\
&&\;\;\;\; + \frac{\hbar}{2} \sum_{\vk \neq 0}
            \int^{\infty}_0  \frac{d\omega}{\pi}
  \ln \left\{ 1 - \Pi_{00}(k, \omega)
          [ W(k) + \Pi_{11}(k, \omega) ]  \right\}  \nonumber \\
&& = \sum_{\vk} \; n_F(\epsilon_{k}-\mu) \; \epsilon_{k}
 + \sum_{\vk \neq 0} \; \frac{1}{2} \hbar \omega_c \;\;
   \hat e(\frac{k}{2k_F}, \frac{\hbar \omega_c}{2\mu} )
\label{ERPA}
\eea
($\hbar \omega_c =\hbar^2/(m_b l_B^2)$).
Here, the first term is the energy of the free CS fermions and
the second term is the contribution of the magneto--plasmon oscillators.
While the first sum is confined to the Fermi sphere, one has to
discuss the convergence of the second sum which we will do below.
In order to simplify the following analysis, we have defined
\be
  \hat e(u,v) =
     - \frac{1}{4} \left( \frac{\tilde \phi}{2} \frac{1}{u^2}
                                 + \frac{\lambda}{uv} \right) + e(u,v) \; ,
\ee
with
\bea
 e(u,v) = \frac{2u^2}{v} \int^{\infty}_0 \frac{dx}{\pi} \;
         \ln \{ 1 + I(x,u,v) \}
\label{euv}
\eea
and
\bea
I(x,u,v) = (1-z_+) &&\left( \frac{\tilde \phi}{2} \frac{v}{2u^2}
        + \frac{\lambda}{2u}
        + (\frac{\tilde \phi}{2})^2 \frac{1}{12} (1-z_+) \right.
\nonumber \\
&&\; \left. \left[(1-z_+)^2 (1-3z_-^2) - \frac{3}{u^2} \right]  \right) .
\eea
From now on, we will use $\lambda = \epsilon^2 k_F /(2\mu)$ as a dimensionless
measure of the Coulomb interaction.
$z_+(x,u)$ and $z_-(x,u)$ are related to the real and imaginary parts of
$z(x,u)$ in (\ref{z}):
\bea
&& z_{\pm}(x,u) =
\nonumber \\
&& \sqrt{\frac{1}{2}
   \left[\sqrt{(1-u^{-2}-x^2)^2+4x^2 } \pm (1-u^{-2}-x^2) \right] } \; .
\eea
The difference between the energy distributions $\hat e$ and $e$
is due to the following:
For large $\omega$, $\Pi_{00}(k, \omega)$ and $\Pi_{11}(k, \omega)$
decay as $\omega ^{-1}$.
Therefore, one has to study the convergence of the $\omega$ integral
in (\ref{ERPA}).
Here, in the first term of an expansion of the $\ln \{\cdots \}$ around $1$,
the factors $\exp(-i\omega 0^+)$ in $\Pi_{00}(k, \omega)$ define the
integration; in the higher terms in the expansion, these convergence
factors can be put equal to one.
$e(u,v)$ is the expression which one would obtain setting these factors
equal to one everywhere.
The difference between $\hat e$ and $e$ thus comes from the
difference in the first term in the expansion of the $\ln \{\cdots \}$,
calculated once with and once without convergence factors.

Now, we analyse the contribution of the magneto--plasmon oscillators
to $E_{\mbox{\scriptsize RPA}}$ by studying the dependence of $\hat e(u,v)$ on $u=k/(2k_F)$.

\begin{figure}[htb]
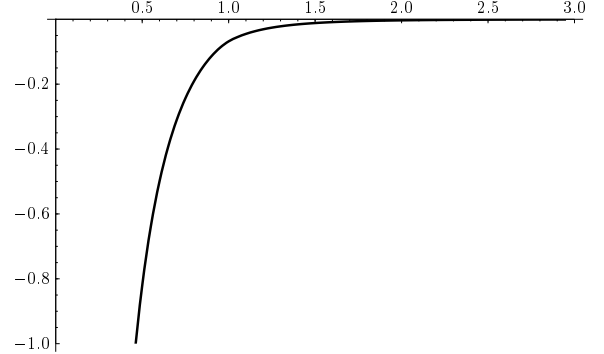

\setlength{\epsfxsize}{0.9\columnwidth}
\centerline{\Bildzwei}
\caption{$\hat e(u,1)$ for $\tilde \phi =2$ and $\lambda =0.7$.}
\label{ehutnum}
\end{figure}

(i) We first discuss the behavior of $\hat e(u,v)$ for large $u$.
With
\bea
 z_+(x,u) &=& 1 - \frac{1}{2u^2} \frac{1}{1+x^2}
              - \frac{1}{8u^4} \frac{1-3x^2}{(1+x^2)^3}
              + {\cal O}(u^{-6}) \;, \nonumber \\
 z_-(x,u) &=& x + {\cal O}(u^{-2}) \;
\eea
a straight forward expansion of $e(u,v)$ yields
\be
 \hat e(u,v) = - \frac{1}{2v} \frac{1}{(2u)^4}
 \left[ (\frac{\tilde \phi}{2})^2 +  \frac{1}{2} \lambda ^2 \right]
 + {\cal O}(u^{-5}) \; .
\ee
Thus, the integral defining $E_{\mbox{\scriptsize RPA}}$ is
{\it convergent for large wave vectors $k$}.
There is no need of an artificial cut off in the ultraviolet.
A numerical integration of (\ref{euv}) shows in fact that the cut off
is rather abrupt, cf. Fig.~\ref{ehutnum}.
We shall justify below the values $v=1$ and $\lambda= 0.7$, which we have chosen here
for $v$ and the strength of the Coulomb interaction.

\begin{figure}[htb]
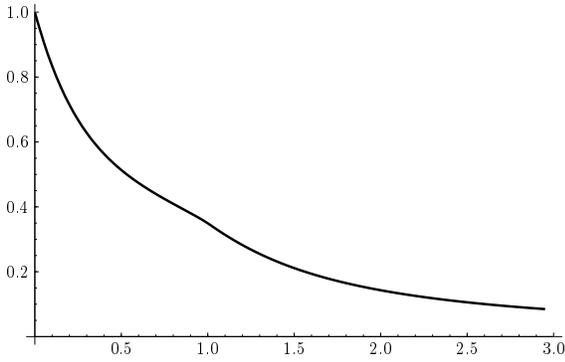

\setlength{\epsfxsize}{0.9\columnwidth}
\centerline{\Bilddrei}
\caption{$e(u,1)$ for $\tilde \phi =2$ and $\lambda =0.7$.}
\label{enume}
\end{figure}

(ii) The opposite case of small $u$ leads to a divergence.
The integral (\ref{euv}) is finite; its dominant contributions come from
$x \propto u^{-2}$. We find for $u=0$
\be
  e(0,v) = \sqrt{\frac{1}{v} \frac{\tilde \phi}{2} } \; .
\ee
The result of the numerical integration for general $u$ as shown
in Fig.~\ref{enume} confirms a smooth behavior of $e(u,v)$ for
small $u$.
The negative curvature in the region near $u=0.9$ is seen to be related
to the value of $\lambda$. This feature gets less pronounced as $\lambda$
is decreased.
Hence, $\hat e(u,v) \propto -\tilde \phi /(8u^2) \propto k^{-2}$ for small $k$
and $E_{\mbox{\scriptsize RPA}}$  contains {\it a logarithmic divergence in the
infrared regime}.
It is clear from the derivation, that this logarithmic divergence
comes from the first diagram in the RPA series in (\ref{ring}):
\begin{figure}[htb]
\setlength{\epsfxsize}{0.3\columnwidth}
\centerline{\Bildvier}
\label{bild4}
\end{figure}
\vspace{-0.5cm}
\bea
\label{RPA1}
    &=& \ -\beta\Omega_{\mbox{\scriptsize RPA}}^{(1)} \nonumber \\
    &\sim& \ \frac{\beta}{2F} \sum_{\vk \neq 0} K(k)
     \sum_{\vk_1} n_F(\vk_1 + \vk) n_F(\vk_1)  \nonumber \\
    &\sim& N \frac{1}{4 v} {\tilde \phi}^2 \; \beta \hbar \omega_c \;
            \ln \frac{\kappa}{\eta}  \; .
\eea
Here, we give only the term with the leading singularity for
decreasing infrared cut off $\eta$;
$\kappa \ll k_F$ is an upper wave vector cut off.
The result (\ref{RPA1}) is quickly reproduced if one uses only the
leading term $\hat e(u,v) \sim -\tilde \phi /(8u^2) $ in the integral
for the energy (\ref{ERPA}).
The logarithmic divergence is explained as the result of the long--range
behavior of the CS interaction $K(k)$ in a two--dimensional system.
Thus we see, that within the RPA, the ground--state energy is
infinite.

We wish to add two remarks.
First, this divergence in the energy and the divergence in the
self energy discovered in Ref.~\cite{HLR93} can not have the same origin,
since, there, the divergence depends on the form of the
electron--electron interaction, while here, it is already present in the
absence of a Coulomb interaction and rather a result of the pure CS
interaction.
Second, in other cases of RPA calculations, such a divergence in the energy
does not appear; in the case of the Coulomb interaction in three dimensions,
for example, one has a convergent integral ($\int d^3 k / k^2$)
and the same is true for the Coulomb interaction in two dimensions.
Consequently, for the case of the CS interaction, it becomes necessary
to study diagrams {\it beyond} the RPA.

\subsection{Chemical potential}
The chemical potential, as the ground--state energy, should be derived
from the grand--canonical potential.
Calculating the mean particle number $N$ from $\Omega_{\mbox{\scriptsize RPA}}$
and using its value given by the filling factor (\ref{filling}),
we get at zero temperature
\be
v = \frac{\tilde \phi}{2} \; + 2 {\tilde \phi} v
    \int_0^{\infty} \!du\ u
    \left[ u \partial_u + 2 v \partial_v + \lambda \partial_{\lambda} \right]
    \hat e(u,v) \, .
\label{muee}
\ee
This equation should determine the chemical potential $\mu$ via
$\mu = \frac{1}{2} \hbar \omega_c / v$.
But $\hat e(u,v)$ diverges in the infrared regime as $-\tilde \phi /(8u^2)$.
Therefore, the r.h.s.~of (\ref{muee}) becomes infinitely large
and there is no positive solution for the parameter $v$ from
this equation.
Hence, we have to conclude that within the RPA, there is no
satisfactory solution for the chemical potential.

\subsection{Expansion of the energy in the Coulomb interaction}
Here, we wish to study the expansion of the ground--state energy
in the strength of the Coulomb interaction, $\lambda$.
Generally, we have for the energy per particle
\be
  E/N  = \frac{1}{2} \hbar \omega_c  + E^{(1)}/N + {\cal O}(\lambda^2) \;.
\label{ExpE}
\ee
The first term is the energy of the lowest Landau level; in the RPA,
it diverges as shown above.
Because the CS transformation gives an exact reformulation of the
problem also in the absence of an electron--electron interaction,
this divergence must be compensated by other divergent diagrams;
we expect such a compensation by the diagrams discussed in
Subsec.~\ref{jenseitsRPA} below.
In the calculation of the second term, now, we encounter the problem that the
unperturbed chemical potential can not be determined from the unperturbed
grand--canonical potential because the divergence is not eliminated.
As a way out, we insert in the second term in (\ref{ExpE}) the known
exact unperturbed value $\mu = \hbar \omega_c/2$.
Because for free composite fermions, $\mu = 1/(2m_b) (4\pi N/F)$,
i.e.~$\mu = \hbar \omega_c \nu$ (use eq.(\ref{filling}))
that value of $\mu$ coincides for half--filling accidentally
with the chemical potential of the free composite fermions.
This substitution fixes the variable $v$,  $v=1$.
Using this and taking the electron density
$\rho \approx 1.5 \cdot 10^{15} m^{-2}$
from the experiments of Kang et al. \cite{KSPBW93} in GaAs
($\varepsilon_r = 12.8$ and $m_b = 0.068\; m_{el}$), one derives for
the value of the relative strength of the Coulomb interaction
$\lambda \approx 0.7$ which was used in Figs. 1 and 2 together with $v=1$.
We now obtain for the second term in (\ref{ExpE})
\bea
&&E^{(1)}_{\mbox{\scriptsize RPA}}/N = \nonumber \\
&&\frac{\epsilon^2}{l_B} \frac{2 \tilde \phi}{v^{3/2}}
   \int ^{\infty}_0 du \left\{ -\frac{1}{4} + u^2
   \int ^{\infty}_0 \frac{dx}{\pi}
   \frac{1-z_+(x,u)}{1+I_0(x,u,v)} \right\} \;.
\eea
Here, $I_0 = I\,|_{\lambda = 0}$.
A numerical evaluation is simple and the result is
$E^{(1)}_{\mbox{\scriptsize RPA}}/N = -0.6\; \epsilon^2/l_B$.
This compares reasonably well with
$E^{(1)}_{\mbox{\scriptsize sim}}/N = -0.466\; \epsilon^2/l_B$
obtained in numerical simulations in the spherical geometry
by Morf and d'Ambrumenil \cite{MA95},
and by Girlich \cite{Girlich98} via the threshold energy
of the many--particle density of states.

\subsection{Higher order logarithmic divergences} \label{jenseitsRPA}
In the CS problem at $d=2$, there are logarithmically divergent diagrams
in the grand--canonical potential, because the CS interaction $K(k)$
diverges as $1/k^2$ for small wave vectors.
In contrast, in the Coulomb problems at zero magnetic field in $d=3$ and $d=2$,
such divergences are absent.
The lowest--order divergence is found in the logarithmically divergent
diagram (\ref{RPA1}), the first of the RPA diagrams.
In addition, there are diagrams for $\Omega$ beyond the RPA diverging
with higher powers of $\ln \eta$.
These diagrams are characterized by independent $k_l$--integrations
over $K(k_l)$; they are either ladder type diagrams or maximally
crossed diagrams or combinations of both.
The leading singularities of the ladder diagrams in second
and third order are
\begin{figure}[htb]
\setlength{\epsfxsize}{0.4\columnwidth}
\centerline{\Bildfuenf}
\label{bild5}
\end{figure}
\be
    \ = \ -\beta\Omega_{\mbox{\scriptsize L}}^{(2)}
    \ \sim \  \frac{N}{12v} \; \beta\hbar\omega_c \; {\tilde \phi}^4
    \left( \ln \frac{\kappa}{\eta} \right)^2 \; ,
\ee
\begin{figure}[htb]
\setlength{\epsfxsize}{0.4\columnwidth}
\centerline{\Bildsechs}
\label{bild6}
\end{figure}
\be
    \ = \ -\beta\Omega_{\mbox{\scriptsize L}}^{(3)}
    \ \sim \  \frac{N}{160v^2} \; (\beta\hbar\omega_c)^2 \; {\tilde \phi}^6
    \left( \ln \frac{\kappa}{\eta} \right)^3 \; .
\ee
Similarly, for the maximally crossed diagrams
\begin{figure}[htb]
\setlength{\epsfxsize}{0.4\columnwidth}
\centerline{\Bildsieben}
\label{bild7}
\end{figure}
\be
    \ = \ -\beta\Omega_{\mbox{\scriptsize Lx}}^{(2)}
    \ \sim \ - \frac{N}{48v} \; \beta\hbar\omega_c \; {\tilde \phi}^4
    \left( \ln \frac{\kappa}{\eta} \right)^2 \; ,
\ee
\begin{figure}[htb]
\setlength{\epsfxsize}{0.4\columnwidth}
\centerline{\Bildacht}
\label{bild8}
\end{figure}
\be
    \ = \ - \beta\Omega_{\mbox{\scriptsize Lx}}^{(3)}
    \ \sim \  \frac{N}{144v^2} \; (\beta\hbar\omega_c)^2 \; {\tilde \phi}^6
    \left( \ln \frac{\kappa}{\eta} \right)^3 \; .
\ee
The summation of all logarithmically divergent diagrams 
is expected to give a finite result.
This should then lead to a finite ground state energy and a well defined
equation for the chemical potential in the absence of 
an electron--electron interaction.
Also, such a summation should improve the result for the Coulomb
contribution to the ground--state energy.

\section{Conclusion}
On the basis of bare response functions calculated analytically for
arbitrary $k$ and $\omega$,
we evaluated the grand--canonical potential in the RPA without any
further approximation.
Because of the $1/k^2$--singularity of the Chern--Simons interaction
and the systems dimension $d=2$, the grand--canonical potential contains
a logarithmically divergent term.
Thus, the ground--state energy and the chemical potential
are not well defined within the RPA.
The CS transformation leads to the picture of composite fermions
forming a Fermi liquid at zero magnetic field, which is experimentally
well supported.
However, for a system of non--interacting electrons in a strong magnetic
field, which we call the ``unperturbed problem'',
this transformation already leads to a very complicated reformulation
of the Hamiltonian.
We have to conclude that, in the CS formulation, one has to go
beyond the RPA in order to get a satisfying solution of
that ``unperturbed problem''.

Further, we calculated the first--order term in the expansion of the
ground--state energy with respect to the strength of the Coulomb interaction.
There, we used for the chemical potential, which remains undefined
in the RPA, the unperturbed value.
The result obtained is $\approx $ 25\% larger than the result of
numerical simulations.
In view of the simplicity of the random--phase approximation and
the absence of a small parameter in the ``unperturbed problem'',
this seems to be a reasonable agreement.
For an improvement of the calculation of the Coulomb contribution
to the energy, the ``unperturbed problem'' must be solved in a better
approximation, which should be free from a divergence.

The ``unperturbed problem'' is well defined, the CS transformation is
a rigorous reformulation; therefore, one should not encounter divergences.
Thus, the divergent diagram in the RPA must be compensated by others.
We finally studied other logarithmically divergent diagrams,
beyond the RPA, and calculated their leading singularities.
It is an open question, whether taking these leading singularities into
account is already sufficient for a compensation.
We plan to return to this question in a following publication.

\end{multicols}

\begin{thebibliography}{10}

\bibitem{KSPBW93}
W. Kang {\it et~al.}, Phys. Rev. Lett. {\bf 71},  3850  (1993).

\bibitem{SWBLvKFKGW96}
J.~H. Smet {\it et~al.}, Phys. Rev. Lett. {\bf 77},  2272  (1996).

\bibitem{SDKTPBW94}
H.~L. St\"ormer {\it et~al.}, Semicond. Sci. Technol. {\bf 9},  1853  (1994).

\bibitem{Heiman}
{\em High Magnetic Fields in the Physics of Semiconductors}, edited by D.
  Heiman (World Scientific, Singapore, 1995).

\bibitem{S97}
M. Shayegan, Solid State Commun. {\bf 102},  155  (1997).

\bibitem{Das-Sarma-Pinczuk}
{\em Perspectives in Quantum Hall Effects : Novel Quantum Liquids in
  Low-Dimensional Semiconductor Structures}, edited by S. {Das Sarma} and A.
  Pinczuk (John Wiley \& Sons, New York, 1996).

\bibitem{HLR93}
B.~I. Halperin, P.~A. Lee, and N. Read, Phys. Rev. B {\bf 47},  7312  (1993).

\bibitem{SH95}
A. Stern and B.~I. Halperin, Phys. Rev. B {\bf 52},  5890  (1995).

\bibitem{SSH96}
S.~H. Simon, A. Stern, and B.~I. Halperin, Phys. Rev. B {\bf 54},  R11114
  (1996).

\bibitem{S96}
S.~H. Simon, J. Phys.: Condens. Matter. {\bf 48},  10127  (1996).

\bibitem{KC97}
P. Kopietz and G.~E. Castilla, Phys. Rev. Lett. {\bf 78},  314  (1997).

\bibitem{SM97}
R. Shankar and G. Murthy, Phys. Rev. Lett. {\bf 79},  4437  (1997).

\bibitem{WDK97}
W. Weller, J. Dietel, and T. Koschny, preprint (NTZ 46/1997), U Leipzig
  (unpublished).

\bibitem{MA95}
R. Morf and N. d'Ambrumenil, Phys. Rev. Lett. {\bf 74},  5116  (1995).

\bibitem{Girlich98}
U. Girlich, Ph.D. thesis, U Leipzig (unpublished).

\bibitem{Mahan}
G.~D. Mahan, {\em Many--Particle Physics} (Plenum Press, New York, 1990).

\bibitem{L54}
J. Lindhard, Kgl. Dan. Vidensk. Selsk. Mat.--fys. Medd. {\bf 28},  8  (1954).

\end{thebibliography}
\end{document}